\documentclass[12pt]{iopart}
\usepackage{graphicx}
\usepackage{amsthm}
\usepackage{iopams}
\usepackage{color}
\usepackage{bbm}

\newcommand{\ket}[1]{\mbox{$ | #1 \rangle $}} 
\newcommand{\bra}[1]{\mbox{$ \langle #1 | $}}

\begin{document}

\title[Non-Poissonian statistics from Poissonian light sources]{Non-Poissonian statistics from Poissonian light sources with application to passive 
decoy state quantum key distribution}

\author{Marcos Curty$^1$, Tobias Moroder$^{2,3}$, Xiongfeng Ma$^2$ and Norbert
  L\"{u}tkenhaus$^{2,3}$}   

\address{$^1$ ETSI Telecomunicaci\'on, Department of Signal Theory and Communications, University of Vigo, E-36310 Vigo (Pontevedra), Spain} 
\address{$^2$ Institute for Quantum Computing, University of Waterloo, Waterloo, ON, N2L 3G1, Canada}
\address{$^3$ Quantum Information Theory Group, Institut f\"ur Theoretische Physik I, and Max Planck Institute for the Science of Light, University of Erlangen-N\"urnberg, 91058 Erlangen, Germany}

\eads{\mailto{mcurty@com.uvigo.es}}

\begin{abstract}
We propose a method to prepare different non-Poissonian signal pulses from sources of 
Poissonian photon number 
distribution using only linear optical elements and threshold photon detectors. This method allows a simple passive 
preparation of decoy states for quantum key distribution. We show that the resulting key rates are comparable to the 
performance of active choices of intensities of Poissonian signals.
\end{abstract}

\maketitle

The main benchmark to compare different quantum key distribution (QKD) systems is their secret key rate over a given distance
\cite{qkd}. For example, it is well known that QKD schemes with single photon sources can provide a key generation rate of linear behavior with the transmission efficiency of the quantum channel. 
Unfortunately, single photon sources are still beyond our 
present experimental capability 
and QKD implementations with phase randomized weak coherent 
pulses (WCP) 
are typically employed.
In this context, it has been recently shown that decoy state QKD with WCP can basically 
reach the same performance as single photon sources  \cite{decoy,decoy1,decoy2}. 
The essential idea 
behind decoy state QKD is quite simple. The sender (Alice)
varies, independently and randomly, 
the mean photon number of each signal state she transmits to the receiver (Bob). 
This is usually performed by using a variable optical attenuator 
together with a random number generator. 
From the measurement results corresponding to 
different intensity settings, the legitimate users can obtain a better estimation of the behavior of the 
quantum channel, which translates into an enhancement of the achievable secret key rate and distance. 
This technique has been successfully implemented in several recent experiments \cite{decoy_e,decoy_e1,decoy_e2}, 
which show the practical feasibility of this method.

While active modulation of the intensity of the pulses suffices to perform decoy state QKD in principle, 
in practice passive preparation might be desirable in some scenarios. 
For instance, in those setups operating at high transmission rates. 
Known passive methods rely on the use of 
a parametric down-conversion source together with a photon detector \cite{mauerer,mauerer1,mauerer2}.
In this Letter we show that phase randomized WCP can also be used for the same
purpose, {\it i.e.}, one does not need a non-linear optics network preparing entangled states. 
Note that the crucial requirement of a passive decoy state setup is to have correlations 
between the photon number statistics of different signals; hence it is sufficient that these correlations are classical. 
Our method uses only linear optical elements and a threshold photon detector.
For simplicity, we consider that this detector has perfect detection efficiency and no dark counts.
But this analysis can also be adapted to cover the case of imperfect detectors. 
A similar technique can also be applied to heralded single-photon sources 
showing non-Poissonian photon 
number statistics \cite{masato}.

\begin{figure}
\begin{center}
\includegraphics[angle=0,scale=0.62]{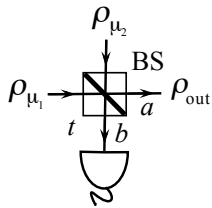}
\end{center}
\caption{Interference of two phase randomized WCP, $\rho_{\mu_1}$ and $\rho_{\mu_2}$, at a beam splitter (BS) 
of transmittance $t$. $a$ and $b$ represent the two output modes.
\label{figure1}}
\end{figure}

The key idea is rather simple, although it is counter-intuitive. It is illustrated in Fig.~\ref{figure1}. 
When
two phase randomized WCP, 
$\rho_{\mu_1}=e^{-\mu_1}\sum_{n=0}^{\infty}\mu_1^n/n!\ket{n}\bra{n}$ and $\rho_{\mu_2}=e^{-\mu_2}\sum_{n=0}^{\infty}\mu_2^n/n!\ket{n}\bra{n}$,
interfere at a beam splitter (BS) of transmittance $t$, the photon number 
statistics of the two outcome signals are classically correlated. 
To see this, let us first consider the interference of two pure coherent states with fixed phase relation, $\ket{\sqrt{\mu_1}e^{i\phi_1}}$ and $\ket{\sqrt{\mu_2}e^{i\phi_2}}$, at a BS. The output signals are given by
$\ket{\sqrt{\mu_1{}t}e^{i\phi_1}+i\sqrt{\mu_2(1-t)}e^{i\phi_2}}_a\ket{i\sqrt{\mu_1{}(1-t)}e^{i\phi_1}+\sqrt{\mu_2{}t}e^{i\phi_2}}_b$. The joint probability $p_{n,m}$
of having $n$ photons in mode $a$ and $m$ 
photons in mode $b$ is the 
product of two Poissonian distributions:
$p_{n,m}=
e^{-\upsilon/2}(\upsilon\gamma)^n/n!\times{}e^{-\upsilon/2}[\upsilon(1-\gamma)]^m/m!$, 
with $\upsilon=\mu_1+\mu_2$,
$\gamma=[\mu_1{}t+\mu_2{}(1-t)+\xi\cos{\theta}]/\upsilon$, $\xi=2\sqrt{\mu_1\mu_2{}(1-t)t}$ and
$\theta=\pi/2+\phi_2-\phi_1$. The case of two phase randomized WCP can be solved by just integrating $p_{n,m}$
over all angles $\theta$, {\it i.e.},
\begin{equation}
p_{n,m}=\frac{\upsilon^{n+m}e^{-\upsilon}}{n!m!}\frac{1}{2\pi}\int_0^{2\pi}\gamma^n(1-\gamma)^m d\theta.
\end{equation}
 
\begin{figure}
\begin{center}
\includegraphics[angle=0,scale=0.72]{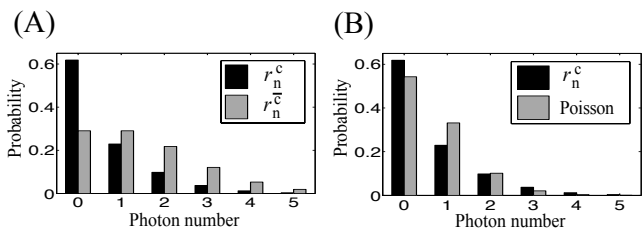}
\end{center}
\caption{
(A) Conditional photon number distribution in mode $a$ (see Fig.~\ref{figure1}): 
$r^{c}_{n}$ (black) and $r^{\bar{c}}_{n}$ (grey) for the case $\mu_1=\mu_2=1$ and $t=1/2$.
(B) $r^{c}_{n}$ (black) versus a Poissonian distribution of the same mean photon number (grey). \label{figure2}}
\end{figure}

By measuring one outcome signal, 
the conditional photon number statistics of the remaining signal
varies depending on the result obtained. Specifically, whenever one ignores the result of the measurement
the total probability of finding  
$n$ photons in mode $a$ can be expressed as 
\begin{equation}\label{help}
p^t_{n}=\sum_{m=0}^{\infty}p_{n,m}=\frac{\upsilon^{n}}{n!}\frac{1}{2\pi}\int_0^{2\pi}\gamma^ne^{-\upsilon{}\gamma} d\theta,
\end{equation}  
which turns out to be a non-Poissonian probability distribution. The joint probability for seeing $n$
photons in mode $a$ and no click in the threshold detector has the form
\begin{equation}\label{pnc}
p^{\bar{c}}_n=p_{n,0}=\frac{\upsilon^{n}e^{-\upsilon}}{n!}\frac{1}{2\pi}\int_0^{2\pi}\gamma^n d\theta.
\end{equation}
If the detector produces a click, 
the joint probability of finding $n$ photons in mode $a$ 
is given by $p^{c}_{n}=p^{t}_{n}-p^{\bar{c}}_{n}$.
Fig.~\ref{figure2} (Case A) shows the conditional photon number statistics
of the outcome signal in mode $a$ depending on the result of the detector
(click and not click): $r^{c}_{n}\equiv{}p^{c}_{n}/(1-F)$ and $r^{\bar{c}}_n\equiv{}p^{\bar{c}}_n/F$, with
$F\equiv{}\sum_{n=0}^{\infty}p^{\bar{c}}_{n}=e^{-[\mu_1{}(1-t)+\mu_2{}t]}I_{0,\xi}$, and where 
$I_{q,z}$ represents the modified Bessel function of the first kind
\cite{Bessel}. This figure
includes as well a comparison between $r^{c}_{n}$ and a 
Poissonian distribution of the same mean photon number (Case B). Both distributions, 
$r^{\bar{c}}_{n}$ and $r^{c}_{n}$, are also non-Poissonian.

To perform decoy state QKD, we consider that Alice and Bob treat no click and click events
separately, and they distill secret key from both of them. We use the secret key rate formula provided 
by Refs.~\cite{gllp,gllp1},
\begin{equation}\label{key_rate}
R\geq{}\textrm{max}\{R^{c},0\}+\textrm{max}\{R^{\bar{c}},0\},
\end{equation}
with
$R^{c}=q\{-Q^{c}f(E^{c})H(E^{c})+p^{c}_{1}Y_1[1-H(e_1)]+p^{c}_{0}Y_0\}$,
and similarly for $R^{\bar{c}}$. The parameter $q$ is the efficiency of the protocol ($q=1/2$ for the standard Bennett-Brassard 1984
protocol \cite{bb84}, and $q\approx{}1$ for its efficient version \cite{eff_bb84}), 
$Q^{c}$ is the overall gain of the signals, 
$E^{c}$ represents the overall quantum bit error rate (QBER), $f(E^{c})$ is the error correction efficiency 
[typically $f(E^{c})\geq{}1$ with Shannon limit $f(E^{c})=1$],   
$Y_n$ denotes the yield of a $n$-photon signal, 
{\it i.e.}, the conditional 
probability of a detection event on Bob's side given that Alice transmits an $n$-photon state,
$e_1$ is the single photon error rate, 
and 
$H(x)=-x\log_2{(x)}-(1-x)\log_2{(1-x)}$ is the binary 
Shannon entropy function. 

The quantities $Q^{c}$, $E^{c}$, $Q^{\bar{c}}$, and $E^{\bar{c}}$ are directly accessible from the experiment. They can be written as
$Q^{c}=\sum_{n=0}^\infty p^{c}_{n}Y_n$ and $Q^{c}E^{c}=\sum_{n=0}^\infty p^{c}_{n}Y_ne_n$, 
and similarly for the case of a no click event. Here 
$e_n$ denotes the error rate of a $n$-photon signal
($e_0=1/2$ for random background).
To apply the secret key rate formula given by Eq.~(\ref{key_rate}) one needs
to estimate a lower bound on $Y_1$, together with an upper 
bound on $e_1$. For that, we follow
the procedure proposed in Ref.~\cite{estimation}. 
This method requires that $p^{t}_{n}$ and $p^{\bar{c}}_{n}$ 
satisfy certain conditions that we checked numerically.
Note, however, that many other 
estimation techniques are also 
available, like, for instance, linear programming tools.  
We obtain
\begin{equation}\label{eq1}
Y_1\geq{}Y^l_1\equiv\textrm{max}\Bigg\{0,\frac{p^{\bar{c}}_2Q^t-p^t_2Q^{\bar{c}} - (p^{\bar{c}}_2p^t_0-p^t_2p^{\bar{c}}_0)Y_0^u} {p^{\bar{c}}_2p^t_1-p^t_2p^{\bar{c}}_1}\Bigg\},
\end{equation}
where $Q^{t}=Q^{c}+Q^{\bar{c}}$, and
$Y_0^u$ denotes an upper bound on the background rate $Y_0$ given by
$Y_0^u=\textrm{min}\{(2E^{\bar{c}}Q^{\bar{c}})/p^{\bar{c}}_0,(2E^{t}Q^{t})/p^{t}_0\}$. The error rate $e_1$ can be 
upper bounded as
\begin{equation}\label{eq2}
e_1\leq{}e^u_1\equiv\textrm{min}\Bigg\{\frac{E^{\bar{c}}Q^{\bar{c}}-p^{\bar{c}}_0Y_0^le_0}{p^{\bar{c}}_1Y_1^l}, \frac{E^{c}Q^{c}-p^{c}_0Y_0^le_0}
{p^{c}_1Y_1^l}, \frac{p^{\bar{c}}_0E^tQ^t
 -p^t_0E^{\bar{c}}Q^{\bar{c}}} 
 {(p^{\bar{c}}_0p^t_1-p^t_0p^{\bar{c}}_1)Y_1^l}
 \Bigg\},
\end{equation} 
with $Q^{t}E^{t}=Q^{c}E^{c}+Q^{\bar{c}}E^{\bar{c}}$, and where 
$Y_0^l$ denotes a lower bound on $Y_0$ given by 
$Y_0\geq{}Y^l_0\equiv\textrm{max}\{0,(p^t_1Q^{\bar{c}}-p^{\bar{c}}_1Q^t)/(p^t_1p^{\bar{c}}_0-p^{\bar{c}}_1p^t_0)\}$.

The only relevant statistics to evaluate $Y^l_0$, $Y^l_1$, and $e^u_1$ are 
$p^{t}_n$ and $p^{\bar{c}}_n$, with $n=0,1,2$. These probabilities can be obtained by solving 
Eqs.~(\ref{help})-(\ref{pnc}). After a short calculation, we find that 
$p^{\bar{c}}_0=e^{-\upsilon}$, $p^{\bar{c}}_1=\omega{}e^{-\upsilon}$, and $p^{\bar{c}}_2=(2\omega^2+\xi^2)e^{-\upsilon}/4$, with 
$\omega=\mu_1{}t+\mu_2(1-t)$.
The probabilities 
$p^{t}_n$ have the form 
$p^{t}_0=I_{0,\xi}e^{-\omega}$, $p^{t}_1=[\omega{}I_{0,\xi}-\xi{}I_{1,\xi}]e^{-\omega}$, 
and $p^{t}_2=[\omega^2I_{0,\xi}+(1-2\omega)\xi{}I_{1,\xi}+\xi^2{}I_{2,\xi}]e^{-\omega}/2$.

For simulation purposes 
we consider the channel model 
used in Refs.~\cite{decoy1,estimation}. This model reproduces a normal behaviour 
of the quantum channel, {\it i.e.}, in the absence of eavesdropping.
It allows us to calculate the observed experimental parameters 
$Q^{\bar{c}}$, $E^{\bar{c}}$, $Q^{t}$, and $E^{t}$. Our
results, however, can also be
applied to any other quantum channel, as they only depend on the observed
gains and QBERs. In the scenario considered, the 
yields have the form 
$Y_n=1-(1-Y_0)(1-\eta)^n$,
where $\eta$ represents the overall 
transmittance of the system \cite{decoy1,estimation}. 
This parameter can be related with a transmission distance $l$ measured in km for 
the given QKD scheme
as $\eta=10^{-\frac{\alpha{}l}{10}}$, where $\alpha$ represents 
the loss coefficient of the optical fiber measured in dB/km. 
The product $Y_ne_n$ can be expressed as 
$Y_ne_n=Y_0e_0+(Y_n-Y_0)e_d$,
where $e_{d}$ is the probability that a photon hits the wrong detector due to the 
misalignment in the quantum channel and in Bob's detection setup \cite{decoy1,estimation}. After substituting these 
definitions into the gain and QBER formulas we obtain
$Q^{\bar{c}}=F-(1-Y_0)e^{(1-\eta)\omega-\upsilon}I_{0,(1-\eta)\xi}$, 
$Q^{\bar{c}}E^{\bar{c}}=(e_0-e_d)Y_0F+e_dQ^{\bar{c}}$, 
$Q^{t}=1-(1-Y_0)e^{-\eta\omega}I_{0,\eta\xi}$ and 
$Q^{t}E^{t}=(e_0-e_d)Y_0+e_dQ^{t}$.

\begin{figure}
\begin{center}
\includegraphics[angle=0,scale=0.35]{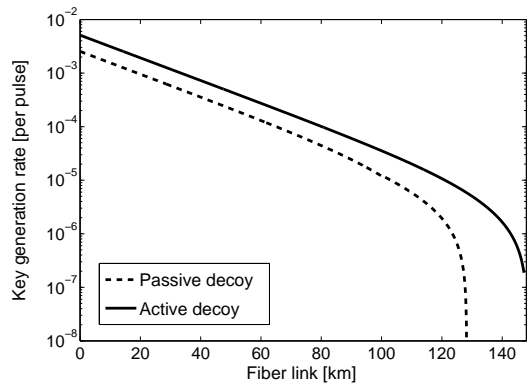}
\end{center}
\caption{
Lower bound on the secret key rate $R$ given by Eq.~(\ref{key_rate}) 
in logarithmic scale for 
a passive decoy state setup with two intensity settings (dashed line). 
The transmittance of the BS is $t=1/2$. 
The solid line represents a lower bound on $R$ for an active asymptotic decoy state 
system \cite{decoy1}. \label{figure3}}
\end{figure}

The resulting secret key rate is illustrated in Fig.~\ref{figure3}. 
The experimental parameters
are \cite{gys}: $Y_0=1.7\times{}10^{-6}$, $e_{d}=0.033$, $\alpha=0.21$ dB/km, and Bob's 
detection efficiency equal to
$0.045$. 
We assume that  
$q=1$, $f(E^{c})=f(E^{\bar{c}})=1.22$,
and $t=1/2$, {\it i.e.}, 
we consider a simple $50:50$ BS.  
With this configuration, it turns out 
that the optimal values of the intensities $\mu_1$ and $\mu_2$ are almost constant with the 
distance. One of them is quite weak (around $10^{-4}$), while the other one is around 
$0.55$. 
Fig.~\ref{figure3} includes as well 
the case of an 
active asymptotic decoy state QKD system \cite{decoy1}. 
The cutoff points where the secret key rate drops down to zero are $l\approx{}128$ km (passive setup
with two intensity settings)
and $l\approx{}147$ km (active 
asymptotic setup). One could reduce this gap further by using 
a passive scheme with more intensity settings.  
For instance, one may employ
a photon number resolving detector instead of a simple threshold photon detector, 
or use more threshold detectors in combination with BS. 
From these results we see that the performance of 
the passive scheme is comparable to the active one, thus showing the practical 
interest of the passive setup.   

To conclude, we have analyzed a simple passive decoy state QKD system with phase randomized WCP. This setup 
represents an alternative to those active schemes based on the use of a variable optical attenuator. 
In the asymptotic limit of an infinite long experiment, we have shown that this passive system
can provide a similar performance to the one achieved with an active source and infinity decoy settings. 
This idea can also be applied to other practical scenarios with different signals and detectors like, for 
example, those based on thermal states or even strong coherent pulses in conjunction with a regular 
photo-detector. Details of this analysis will be presented somewhere else. 

The authors wish to thank R. Kaltenbaek, H.-K. Lo, B. Qi, and
Y. Zhao for very useful discussions. M.C. especially thanks the  
Institute for Quantum Computing (University of Waterloo) for hospitality and support
during his stay in this institution. This work was supported by the European Projects 
SECOQC and QAP, by the NSERC Discovery Grant, Quantum Works, CSEC, and
by Xunta de Galicia (Spain, Grant No. INCITE08PXIB322257PR). 
 
\section*{References}

\bibliographystyle{iopart-num}

\providecommand{\newblock}{}

\end{document}